\documentclass[english,aps,manuscript]{revtex4}
\usepackage[utf8]{inputenc}
\usepackage[a4paper]{geometry}
\usepackage{color}
\usepackage{amsmath}
\usepackage{amssymb}
\usepackage{graphicx}
\usepackage{esint}

\makeatletter
\@ifundefined{textcolor}{}
{%
 \definecolor{BLACK}{gray}{0}
 \definecolor{WHITE}{gray}{1}
 \definecolor{RED}{rgb}{1,0,0}
 \definecolor{GREEN}{rgb}{0,1,0}
 \definecolor{BLUE}{rgb}{0,0,1}
 \definecolor{CYAN}{cmyk}{1,0,0,0}
 \definecolor{MAGENTA}{cmyk}{0,1,0,0}
 \definecolor{YELLOW}{cmyk}{0,0,1,0}
}

\usepackage[english]{babel}

\usepackage{ragged2e}

\makeatletter
\newcommand*{\centerfloat}{%
  \parindent \z@
  \leftskip \z@ \@plus 1fil \@minus \textwidth
  \rightskip\leftskip
  \parfillskip \z@skip}
\makeatother

\makeatother

\usepackage{babel}
\begin{document}

\title{{\Large{}Dynamic information routing in complex networks}}

\author{Christoph Kirst$^{1-5}$, Marc Timme$^{1,3,4}$, and Demian Battaglia$^{4,6}$\vspace{1cm}}

\affiliation{{\footnotesize{}$^{1}$Network Dynamics, Max Planck Institute for
Dynamics and Self-Organization (MPIDS), 37077 Göttingen, Germany,}\\
{\footnotesize{}$^{2}$Nonlinear Dynamics, Max Planck Institute for
Dynamics and Self-Organization (MPIDS), 37077 Göttingen Germany,}\\
{\footnotesize{}$^{3}$Institute for Nonlinear Dynamics, Georg-August
University Göttingen, 37077 Göttingen, Germany,}\\
{\footnotesize{}$^{4}$Bernstein Center for Computational Neuroscience
(BCCN), 37077 Göttingen, Germany, }\\
{\footnotesize{}$^{5}$Center for Physics and Biology, The Rockefeller
University, New York, NY 10065, USA, }\\
{\footnotesize{}$^{6}$Institute of Systems Neuroscience, Aix-Marseille
University, 13005 Marseille, France\vspace{1cm}}}
\begin{abstract}
Flexible information routing fundamentally underlies the function
of many biological and artificial networks. Yet, how such systems
may specifically communicate and dynamically route information is
not well understood. Here we identify a generic mechanism to route
information on top of collective dynamical reference states in complex
networks. Switching between collective dynamics induces flexible reorganization
of information sharing and routing patterns, as quantified by delayed
mutual information and transfer entropy measures between activities
of a network's units. We demonstrate the power of this generic mechanism
specifically for oscillatory dynamics and analyze how individual unit
properties, the network topology and external inputs coact to systematically
organize information routing. For multi-scale, modular architectures,
we resolve routing patterns at all levels. Interestingly, local interventions
within one sub-network may remotely determine non-local network-wide
communication. These results help understanding and designing information
routing patterns across systems where collective dynamics co-occurs
with a communication function.
\end{abstract}
\maketitle
Attuned function of many biological or technological networks relies
on the precise yet dynamic communication between their subsystems.
For instance, the behavior of cells depends on the coordinated information
transfer within gene regulatory networks \cite{Tyson2001,Tkacik2008}
and flexible integration of information is conveyed by the activity
of several neural populations during brain function \cite{Tononi1998}.
Identifying general mechanisms for the routing of information across
complex networks thus constitutes a key theoretical challenge with
applications across fields, from systems biology to the engineering
of smart distributed technology \cite{Weber2005,Stricker2008,Varela2001}. 

Complex systems with a communication function often show characteristic
dynamics, such as oscillatory or synchronous collective dynamics with
a stochastic component \cite{Winfree1980,Kuramoto1984,Strogatz2001,Pikovsky2003,Acebron2005}.
Information is carried in the presence of these dynamics within and
between neural circuits \cite{Fries2005,Salazar2012}, living cells
\cite{Goldbeter2002,Feillet2014}, ecologic or social groups \cite{Blasius1999,Neda2000a}
as well as technical communication systems, such as ad hoc sensor
networks \cite{Kiss2007,Klinglmayr2012}. While such dynamics could
simply reflect the properties of the interacting unit's, emergent
collective dynamical states in biological networks can actually contribute
to the system's function. For example, it has been hypothesized that
the widely observed oscillatory phenomena in biological networks enable
emergent and flexible information routing \cite{Fries2005}.

Here we derive a theory that shows how information if conveyed by
fluctuations around collective dynamical reference states (e.g.~a
stable oscillatory pattern) can be flexibly routed across complex
network topologies. Quantifying information sharing and transfer by
time-delayed mutual information \cite{Shaw1981,Vastano1988} and transfer
entropy \cite{Schreiber2000} curves between time-series of the network's
units, we demonstrate how switching between multi-stable states enables
the rerouting of information without any physical changes to the network.
In fully symmetric networks, anisotropic information transfer can
arise via symmetry breaking of the reference dynamics. For networks
of coupled oscillators our approach gives analytic predictions how
the physical coupling structure, the oscillators' properties and the
dynamical state of the network co-act to produce a specific communication
pattern. Resorting to a collective-phase description \cite{Kawamura2008},
our theory further resolves communication patterns at all levels of
multi-scale, modular topologies \cite{Newman2011,Pajevic2012}, as
ubiquitous, e.\,g.,~in the brain connectome and bio-chemical regulatory
networks \cite{Song2005,Perin2010,Hagmann2008,Barabasi2002}. We thereby
uncover how local interventions within one module may remotely modify
information sharing and transfer between other distant sub-networks.
A combinatorial number of information routing patterns in networks
emerge due to switching between multi-stable dynamical states that
are localized on individual sub-sets of network nodes. 

These results offer a generic mechanism for self-organized and flexible
information routing in complex networked systems. For oscillatory
dynamics the links made between multi-scale connectivity, collective
network dynamics and flexible information routing has potential applications
to the reconstruction and design of gene regulatory circuits \cite{McMillen2002,Feillet2014},
wireless communication networks \cite{Weber2005,Klinglmayr2012} or
to the analysis of cognitive functions \cite{Koepsell2008,Cabral2011,Kopell2014,Belitski2008,ArgwalSommer},
among others.

\section{Information Routing via Collective Dynamics}

To understand how bits of information from external or locally computed
signals can be specifically distributed through a network or to it's
downstream components we first consider a generic stochastic dynamical
system that evolves in time $t$ according to 
\begin{equation}
\frac{d}{dt}x=f\left(x\right)+\xi\label{eq: general SDE}
\end{equation}
where $x=\left(x_{1},\dots,x_{N}\right)$ denotes the variables of
the network nodes, $f$ describes the intrinsic dynamics of the network,
and $\xi=$$\left(\xi_{1},\dots,\xi_{N}\right)$ is a stochastic external
input driving instantaneous state variable fluctuations which carry
the information to be routed through the network. We consider a deterministic
reference state $x^{\left(\mathrm{ref}\right)}\left(t\right)$ solving
\eqref{eq: general SDE} in the absence of signals ($\xi=0$).

To quantify how bits of information 'surfing' on top of such a dynamical
state are routed through the network, we use information theoretic
measures that quantify the amount of information shared and transferred
between nodes, independent of how this information is encoded or decoded.
More precisely, we measure information sharing between signal $x_{i}\left(t\right)$
and the time $d$ lagged signal $x_{j}\left(t+d\right)$ of nodes
$i$ and $j$ in the network via \textit{\emph{the time-delayed mutual
information (dMI)}} \cite{Shaw1981,Vastano1988} 

\begin{equation}
\mathrm{dMI}_{ij}(d)=\iint p_{ij^{\left(d\right)}}\left(t\right)\log\left(\frac{p_{ij^{\left(d\right)}}\left(t\right)}{p_{i}\left(t\right)p_{j}\left(t\right)}\right)\,\mathrm{d}x_{i}\left(t\right)\mathrm{d}x_{j}\left(t+d\right)\label{eq:MI_general}
\end{equation}
Here $p_{i}\left(t\right)$ is the probability distribution of the
variable $x_{i}\left(t\right)$ of unit $i$ at time $t$ and $p_{i,j^{\left(d\right)}}\left(t\right)$
the joint distribution of $x_{i}\left(t\right)$ and the variable
$x_{j}\left(t+d\right)$ lagged by $d$. \textit{\emph{As a second
measure we use the delayed}}\textit{ }\textit{\emph{transfer entropy
(dTE)}} \cite{Schreiber2000} (cf.~Methods) that genuinely measures
information transfer between pairs of units \cite{Lizier2008}. Asymmetries
in the dMI and dTE curves $\mathrm{dMI}_{ij}(d)$ and\textit{\emph{
$\mathrm{dTE}_{i\rightarrow j}(d)$}} then indicate the dominant direction
in which information is shared or transferred between nodes.

To identify the role of the underlying reference dynamical state $x^{\left(\mathrm{ref}\right)}\left(t\right)$
for network communication a small noise expansion in the signals $\xi$
turns a out to be ideally suited: while the small noise expansion
limits the analysis to the vicinity of a specific reference state
which is usually regarded as a weakness, in the context of our study,
this property is highly advantageous as it directly conditions the
calculations on a particular reference state and enables us to extract
it's role for the emergent pattern of information routing within the
network. For white noise sources $\xi$ this method yields general
expressions for the conditional probabilities $p\left(x\left(t+d\right)|x\left(t\right)\right)$
that depend on $x^{\left(\mathrm{ref}\right)}\left(t\right)$. Using
this result the expressions for the delayed mutual information \eqref{eq:MI_general}
and transfer entropy \eqref{eq:TE_general} $\mathrm{dMI}_{i,j}\left(d\right)$
and $\mathrm{dTE}_{i\rightarrow j}\left(d\right)$ become a function
of the underlying collective reference dynamical state (cf.\ Methods
and Supplementary Section 1). The dependency on this reference state
then provides a generic mechanism to change communication in networks
by manipulation the underlying collective dynamics. In the following
we show how this general principle gives rise to a variety of mechanisms
to flexibly change information routing in networks. We focus on oscillatory
phenomena widely observed in networks with a communication function
\cite{Koepsell2008,Belitski2008,ArgwalSommer,Lauschke,Hoppensteadt2000}.

\section{Information exchange in phase signals}

Oscillatory synchronization and phase locking \cite{Kuramoto1984,Pikovsky2003}
provide a natural way for the temporal coordination between communicating
units. Key variables in oscillator systems are the phases $\phi_{i}\left(t\right)$
at time $t$ of the individual units $i$. In fact, a wide range of
oscillating systems display similar phase dynamics \cite{Kuramoto1984,Acebron2005}
(cf.\ Supplementary Section 2) and phase-based encoding schemes are
common, e.g.\ in the brain \cite{Koepsell2008,Belitski2008,ArgwalSommer},
genetic circuits \cite{Lauschke} and artificial systems \cite{Hoppensteadt2000}. 

We first focus on systems in a stationary state with a stationary
distribution for which the expressions for the dMI and dTE become
independent of the starting time $t$ and only depend on the lag $d$
and reference state $\phi^{\left(\mathrm{ref}\right)}\left(t\right)$.
To assess the dominant direction of the shared information between
two nodes we quantify asymmetries in the dMI curve by using the difference
$\delta\mathrm{MI}_{i,j}=\mathrm{MI}_{i\rightarrow j}-\mathrm{MI}_{j\rightarrow j}$
between the integrated mutual informations $\mathrm{MI}_{i\rightarrow j}=\int_{0}^{\infty}\mathrm{dMI}_{i,j}\left(\delta\right)\mathrm{d}\delta$
and $\mathrm{MI}_{j\rightarrow i}$. If this is positive, information
is shared predominantly from unit $i$ to $j$ while negative values
indicate the opposite direction. Analogously, we compute the differences
in dTE as $\delta\mathrm{TE}_{i,j}$(cf.~Methods and Supplementary
Section 3). The set of pairs $\left\{ \delta\mathrm{MI}_{i,j}\right\} $
or $\left\{ \delta\mathrm{TE}_{i,j}\right\} $ for all $i$, $j$
then capture strength and directionality of information routing in
the network akin to a functional connectivity analysis in neuroscience
\cite{Friston2011}. We refer to them as information routing patterns
(IRPs).

\begin{figure*}[!tp]
\begin{raggedright}
\textbf{\centerfloat}\includegraphics[scale=0.92]{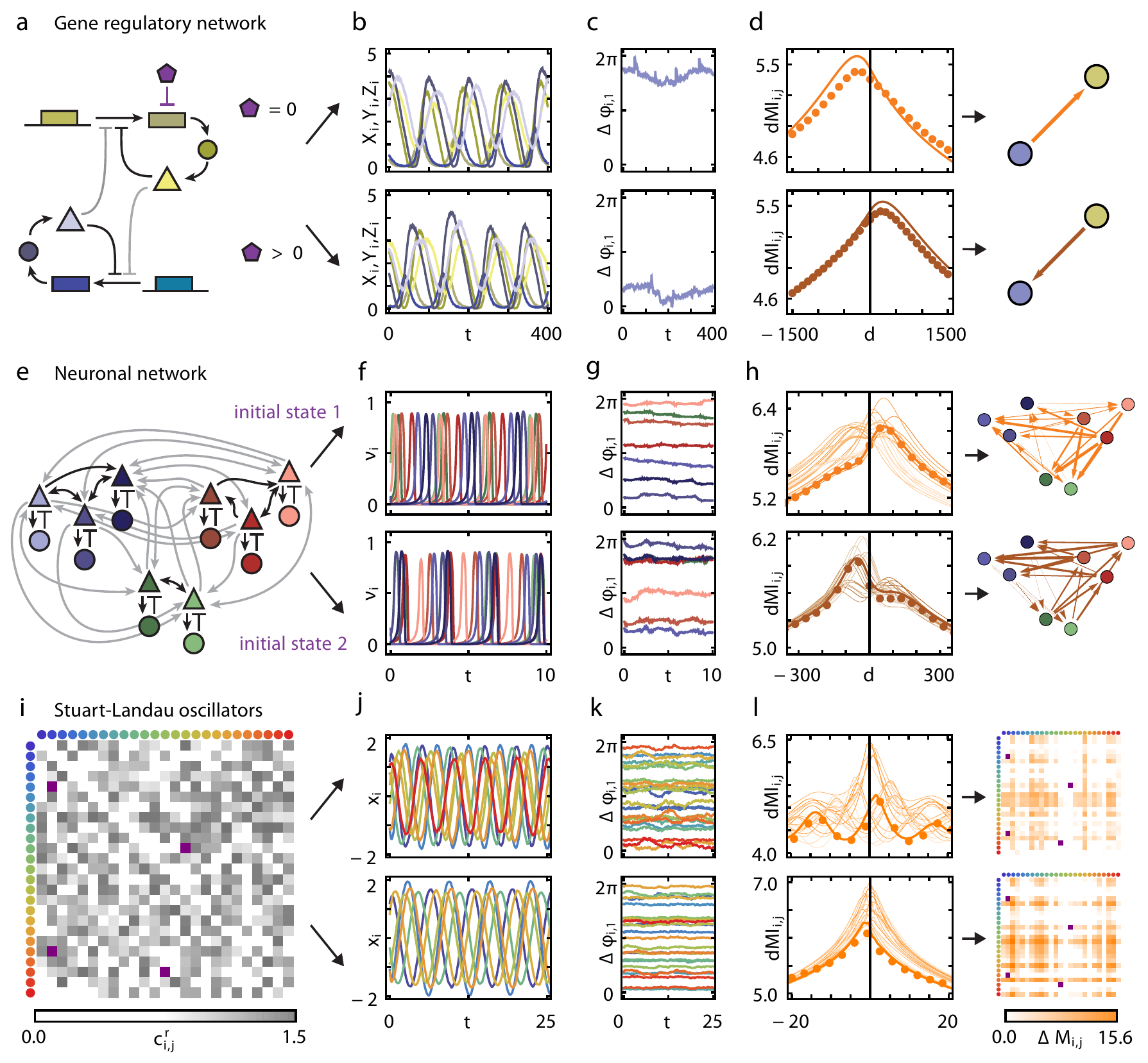}
\par\end{raggedright}

\protect\caption{\label{fig:Example_Networks}\textbf{Flexible information routing
across oscillatory networks}\textbf{\emph{.}} \textbf{a,} Simple model
of a gene regulatory network of two coupled biochemical oscillators
of Goodwin type (yellow and blue). An additional molecule (purple)
degrades the transcribed mRNA in one of the oscillators and thereby
changes its intrinsic frequency. Coupling strengths are gray coded
(darker color indicates stronger coupling), sharp arrows indicate
activating and blunt arrows inhibiting influences. \textbf{b,} Stochastic
oscillatory dynamics of the system's variables.\textbf{ c,} Fluctuations
of the phases extracted from the full dynamics relative to a reference
unit. \textbf{d,} Delayed mutual information ($\mathrm{dMI}_{1,2}$)
between the phase signals. The numerical data (dots) agrees well with
the theoretical prediction \eqref{eq:MI-1} (solid lines). The asymmetry
in the dMI curves around $d=0$ indicates a directed information sharing
pattern summarized in the graphs (right). Arrow thickness indicates
the strength of directed information sharing $\Delta\mathrm{MI}_{i,j}$
measured by the positively rectified differences of the areas below
the integrated $\mathrm{dMI}_{i,j}\left(d\right)$ curve for $d<0$
and $d>0$. \textbf{e--h,} Same as in a--d but for a modular network
of coupled neuronal sub-populations consisting each of excitatory
(triangle) and inhibitory (disk)}
\end{figure*}

\begin{figure}
\justify populations (Wilson-Cowan type dynamics). For the same network
two different collective dynamical states accessed by different initial
conditions give rise to two different information sharing patterns
(f--h top vs. bottom). \textbf{i--l, }As in a--d but for generic oscillators
close to a Hopf bifurcation (Stuart-Landau oscillators) connected
to a larger network. In i and l connectivity matrices are shown instead
of graphs. Two different network-wide information routing patterns
arise (top vs.~bottom in j--l) by changing a small number of connection
weights (purple entries in i and l).\vspace{0.8cm}
\end{figure}

A range of networks of oscillatory units, with disparate physical
interactions, connection topologies and external input signals support
multiple IRPs. For instance, in a model of a gene-regulatory network
with two oscillatory sub-networks (Fig.~\ref{fig:Example_Networks}a)
dMI analysis reveals IRPs with different dominant directions (Fig.~\ref{fig:Example_Networks}b-d,
upper vs.~lower sub-panels). The change is triggered by adding an
external factor that degrades the transcribed mRNA in one of the oscillators
and thereby changes its intrinsic frequency (see Methods). More complex
changes in IRPs emerge in larger networks, possibly with modular architecture.
In a network of interacting neuronal populations (Fig.~\ref{fig:Example_Networks}e)
different initial conditions lead to different underlying collective
dynamical states. Switching between them induces complicated but specific
changes in the IRPs (Fig.~\ref{fig:Example_Networks}f-h). Different
IRPs also emerge by changing a small number of connections in larger
networks. Fig.~\ref{fig:Example_Networks}i-l illustrates this for
a generic system of coupled oscillators each close to a Hopf bifurcation.

In general, several qualitatively different options for modifying
network-wide IRPs exist, all of which are relevant in natural and
artificial systems: (i) changing the intrinsic properties of individual
units (Fig.~\ref{fig:Example_Networks}a-d, cf. also Fig.~\ref{fig: Control}a-c
below), (ii) modifying the system connectivity (Fig.~\ref{fig:Example_Networks}i-l,
Fig.~\ref{fig: Control}d-f) and (iii) selecting distinct dynamical
states of structurally the same system (Fig.~\ref{fig:Example_Networks}e-h,
see also Fig.~\ref{fig:Combinatorics} below).

\section{Theory of Phase Information Routing}

To reveal how different IRPs arise and how they depend on the network
properties and dynamics, we derive analytic expressions for the dMI
and dTE between all pairs of oscillators in a network. We determine
the phase of each oscillator $i$ in isolation by extending its phase
description to the full basin of attraction of the stable limit cycle
\cite{Kuramoto1984,Teramae2009}. For weak coupling, the effective
phase evolution becomes 
\begin{equation}
\frac{\mathrm{d}}{\mathrm{d}t}\phi_{i}=\omega_{i}+\sum_{j=1}^{N}\gamma_{ij}(\phi_{i}-\phi_{j})+\sum_{k=1}^{N}\varsigma_{ik}\xi_{k}\label{eq: dynamical system}
\end{equation}
where $\omega_{i}$ is the intrinsic oscillation frequencies of node
$i$ and the coupling functions $\gamma_{ij}(.)$ depend on the phase
differences only. The final sum in \eqref{eq: dynamical system} models
external signals as independent Gaussian white noise processes $\xi_{k}$
and a covariance matrix $\varsigma_{ik}$. The precise forms of $\gamma_{ij}(.)$
and $\varsigma_{ik}$ generally depend on the specific system (Supplementary
Section 2).

As visible from Fig.~\ref{fig:Example_Networks}e-h, the IRP strongly
depends on the underlying collective dynamical state. We therefore
decompose the dynamics into a deterministic reference part $\phi_{i}^{(\mathrm{ref})}$
and a fluctuating component $\phi_{i}^{(\mathrm{fluct})}$. We focus
on phase-locked configurations for the deterministic dynamics with
constant phase offsets $\Delta\phi_{ij}^{(\mathrm{ref})}=\phi_{i}^{(\mathrm{ref})}-\phi_{j}^{(\mathrm{ref})}$.
We estimate the stochastic part $\phi_{i}^{(\mathrm{fluct})}$ via
a small noise expansion (Methods, Supplementary Theorem~1) yielding
a first-order approximation for the joint probabilities $p_{i,j^{\left(d\right)}}$.
Using \eqref{eq:MI_general} together with the periodicity of the
phase variables, we obtain the delayed mutual information 

\emph{
\begin{equation}
\mathrm{dMI}_{ij}(d)=\frac{k_{ij^{\left(d\right)}}\, I_{1}\left(k_{ij^{\left(d\right)}}\right)}{I_{0}\left(k_{ij^{\left(d\right)}}\right)}-\log\left(I_{0}\left(k_{ij^{\left(d\right)}}\right)\right)\label{eq:MI-1}
\end{equation}
}between phase signals in coupled oscillatory networks; here $I_{n}\left(k\right)$
is the $n^{\mathrm{th}}$ modified Bessel function of the first kind,
and $k_{ij^{(d)}}$ is the inverse variance of a von Mises distributions
ansatz for $p_{ij^{(d)}}$. The system's parameter dependencies,
including different inputs, local unit dynamics, coupling functions
and interaction topologies are contained in $k_{ij^{(d)}}$. By similar
calculations we obtain analytical expressions for $\mathrm{dTE}_{i\rightarrow j}$
(Methods and Supplementary Theorem~2). Our theoretical predictions
well match the numerical estimates (Fig.~\ref{fig:Example_Networks}d,h,l,
see also Fig.~\ref{fig: Mechanisms}c,d below and Supplementary Figs.~1,
2 and 7). For independent input signals ($\varsigma_{ik}=0$ for $i\ne k$)
we typically obtain similar IRPs determined either by the delayed
mutual information or the transfer entropy (Supplementary Fig.~2).

\section{Mechanism of Anisotropic Information Routing}

\begin{figure}[!tph]
\begin{centering}
\includegraphics[scale=0.92]{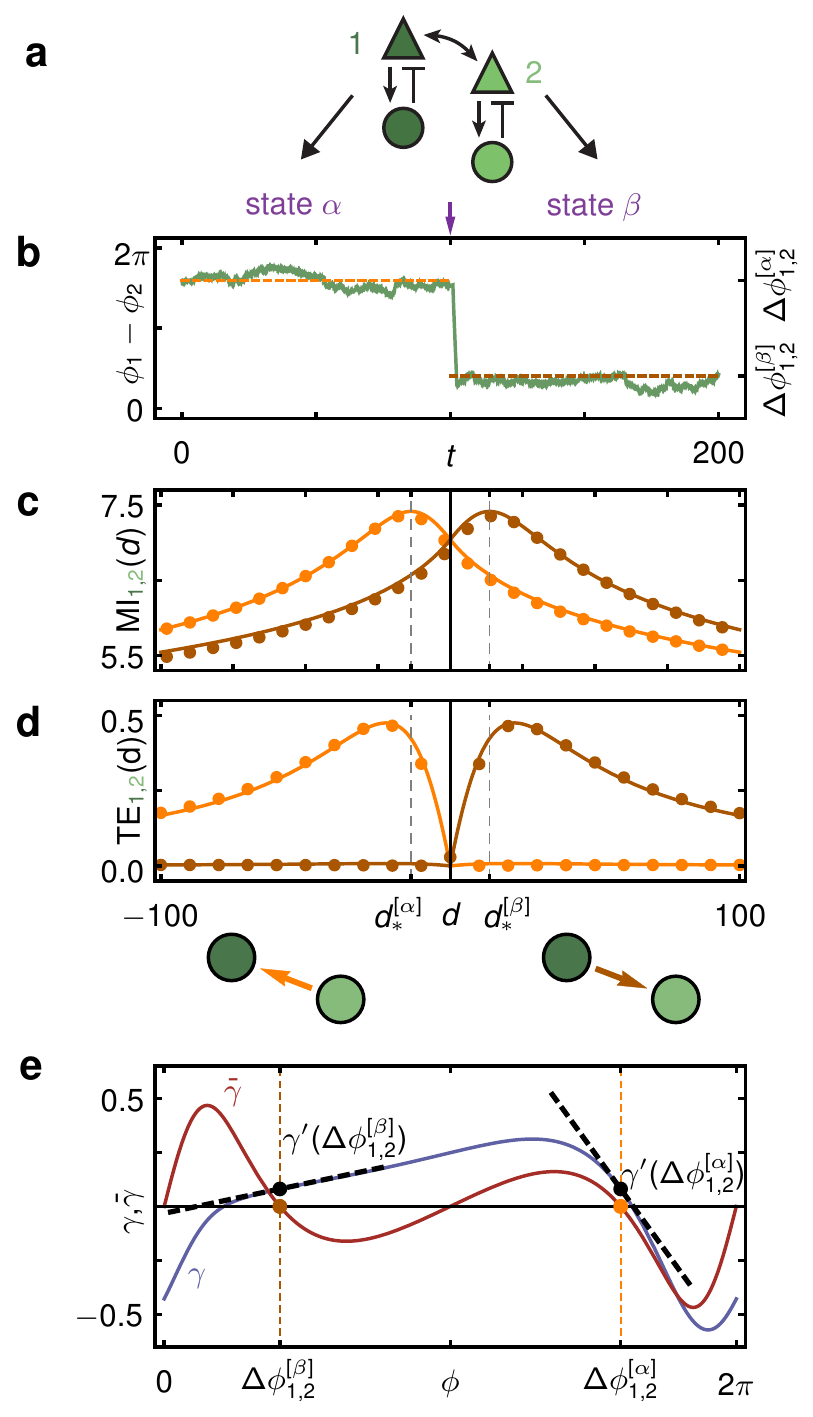}
\par\end{centering}

\protect\caption{\label{fig: Mechanisms}\textbf{Multi-stable dynamics and flexible
anisotropic information routing.}\emph{ }\textbf{a,} Two identical
and symmetrically coupled neuronal circuits of Wilson-Cowan type (dark
and light green, modular sub-network in Fg.~\ref{fig:Example_Networks}e).
\textbf{b,} Phase difference $\Delta\phi_{1,2}(t):=\phi_{1}\left(t\right)-\phi_{2}\left(t\right)$
between the extracted phases of the two neuronal populations fluctuating
around a locked value $\Delta\phi_{1,2}^{\left[\alpha\right]}$ of
a stable deterministic collective state $\alpha$ (orange); a strong
external perturbation (purple arrow) induces a switch to stochastic
dynamics around an alternate stable state $\beta$ (brown) with phase
difference $\Delta\phi_{1,2}^{\left[\beta\right]}$. \textbf{c,} Delayed
mutual information $\mathrm{dMI_{1,2}}$ and \textbf{d, }transfer
entropy $\mathrm{dTE}_{1\rightarrow2}$ between the phase signals
in state $\alpha$ (orange) and $\beta$ (brown) for numerical data
(dots) and theory (lines). The change in peak latencies form $d_{*}^{[\alpha]}<0$
to $d_{*}^{[\beta]}>0$ in the $\mathrm{dMI_{1,2}}$ and the asymmetry
of the $\mathrm{dTE}_{1\rightarrow2}$ curves show anisotropic information
routing. Switching between the two dynamical states reverses the information
flow pattern (graphs, bottom). \textbf{e,} Phase coupling function
$\gamma\left(\Delta\phi\right)=\gamma_{12}\left(\Delta\phi\right)=\gamma_{21}\left(\Delta\phi\right)$
(blue) and its antisymmetric part $\bar{\gamma}\left(\Delta\phi\right)=\gamma\left(\Delta\phi\right)-\gamma\left(-\Delta\phi\right)$
(red). The two zeros of $\bar{\gamma}\left(\Delta\phi\right)$ with
negative slope indicate the deterministic equilibrium phase differences
$\Delta\phi_{1,2}^{\left[\alpha\right]}$ and $\Delta\phi_{1,2}^{\left[\beta\right]}$
in states $\alpha$ and $\beta$, receptively. The directionality
in the information routing pattern arises due to the different slopes
of $\gamma\left(\Delta\phi\right)$ (dashed lines) at the noiseless
phase-locking offsets $\Delta\phi_{1,2}^{\left[\alpha\right]}$ and
$\Delta\phi_{1,2}^{\left[\beta\right]}$.}
\end{figure}

To better understand how a collective state gives rise to a specific
routing pattern with directed information sharing and transfer, consider
a network of two symmetrically coupled identical neural population
models (Fig.~\ref{fig: Mechanisms}a). Due to permutation symmetry,
the coupling functions $\gamma_{ij\,},$ obtained from the phase-reduction
of the original Wilson-Cowan-type equations \cite{Wilson1972} (Methods,
Supplementary Section~5), are identical. For biologically plausible
parameters this network in the noiseless-limit has two stable phase-locked
reference states ($\alpha$ and $\beta$). The fixed phase differences
$\Delta\phi_{12}^{[\alpha]}$ and $\Delta\phi_{12}^{[\beta]}$ are
determined by the zeros of the anti-symmetric coupling $\bar{\gamma}(\Delta\phi)=\gamma(\Delta\phi)-\gamma(-\Delta\phi)$
with negative slope (Fig.~\ref{fig: Mechanisms}e). For a given level
of (sufficiently weak) noise, the system shows fluctuations around
either one of these states (Fig.~\ref{fig: Mechanisms}b) each giving
rise to a different IRP. Sufficiently strong external signals can
trigger state switching and thereby effectively invert the dominant
communication direction visible from the $\mathrm{dMI}$ (Fig.~\ref{fig: Mechanisms}c)
and even more pronounced from the $\mathrm{dTE}$ (Fig.~\ref{fig: Mechanisms}d)
without changing any structural properties of the network.

The anisotropy in information transfer in the fully symmetric network
is due to symmetry broken dynamical states. For independent noise
inputs, $\varsigma_{ik}=\varsigma_{i}\delta_{ik}$, that are moreover
small, the evolution of $\phi_{i}^{(\mathrm{fluct})}$, $i\in\{1,2\}$,
near the reference state $\alpha$ reduces to 
\begin{equation}
\frac{\mathrm{d}}{\mathrm{d}t}\phi_{i}^{(\mathrm{fluct})}=g_{i}^{\left[\alpha\right]}\left(\phi_{i}^{\left(\mathrm{fluct}\right)}-\phi_{j}^{\left(\mathrm{fluct}\right)}\right)+\varsigma_{i}\xi_{i}\label{eq:firstnoisexpK2}
\end{equation}
with coupling constants $g_{1}^{\left[\alpha\right]}=\gamma'(\Delta\phi_{12}^{\left[\alpha\right]})$,
$g_{2}^{\mathbf{\left[\alpha\right]}}=-\gamma'(2\pi-\Delta\phi_{12}^{\left[\alpha\right]})$
(Methods). As $g_{2}^{\left[\alpha\right]}\approx0$ (Fig.~\ref{fig: Mechanisms}e),
the phase $\phi_{2}^{(\mathrm{fluct})}$ essentially freely fluctuates
driven by the noise input $\varsigma_{2}\xi_{2}$. This causes the
system to deviate from the equilibrium phase difference $\Delta\phi_{12}^{\left[\alpha\right]}$.
At the same time, the strongly negative coupling $g_{1}^{\left[\alpha\right]}$
dominates over the noise term $\varsigma_{1}\xi_{1}$ and unit $1$
is driven to restore the phase-difference by reducing $|\phi_{1}^{(\mathrm{fluct})}-\phi_{2}^{(\mathrm{fluct})}|$.
Thus, $\phi_{1}^{(\mathrm{fluct})}$ is effectively enslaved to track
$\phi_{2}^{(\mathrm{fluct})}$ and information is routed from unit
$2$ to unit $1$, reflected in the $\mathrm{dMI}$ and $\mathrm{dTE}$
curves. The same mechanism accounts for the reversed anisotropy in
communication when the system is near state $\beta$ as the roles
of unit $1$ and $2$ are exchanged. Calculating the peak of the $\mathrm{dMI}$
curve in this example also provides a time scale $d_{*}^{[\alpha]}\approx-\log\left(2\right)/g_{1}^{\left[\alpha\right]}$
at which maximal information sharing is observed (Methods, Eq.~\eqref{eq: time scale}).
It furthermore becomes clear that the directionality of the information
transfer in general need not be related to the order in which the
oscillators phase-lock because the phase-advanced oscillator can either
effectively pull the lagging one, or, as in this example, the lagging
oscillator can push the leading one to restore the equilibrium phase-difference.

In summary, effective interactions local in state space and controlled
by the underlying reference state together with the noise characteristics
determine the IRPs of the network. Symmetry broken dynamical states
then induce anisotropic and switchable routing patterns without the
need to change the physical network structure.

\section{Information Routing in Networks of Networks}

For networks with modular interaction topology \cite{Newman2011,Pajevic2012,Song2005,Hagmann2008,Perin2010},
our theory relating topology, collective dynamics and IRPs between
individual units can be generalized to predict routing between entire
modules. Assuming that each sub-network $X$ in the noise-less limit
has a stable phase-locked reference state, a second phase reduction
\cite{Kawamura2008} generalized to stochastic dynamics characterizes
each module by a single meta-oscillator with collective phase $\Phi_{X}$
and frequency $\Omega_{X}$, driven by effective noise sources $\Xi_{X}$
with covariances $\Sigma_{X,Y}$. The collective phase dynamics of
a network with $M$ modules then satisfies 
\begin{equation}
\frac{\mathrm{d}}{\mathrm{d}t}\Phi_{X}=\Omega_{X}+\sum_{Y=1}^{M}\Gamma_{X,Y}(\Phi_{X}-\Phi_{Y})+\sum_{Y=1}^{M}\Sigma_{X,Y}\Xi_{Y}\label{eq: reduced dynamical system}
\end{equation}
where $\Gamma_{X,Y}$ are the effective inter-community couplings
(Supplementary Section 4). The structure of equation~\eqref{eq: reduced dynamical system}
is formally identical to equation~\eqref{eq: dynamical system} so
that the expressions for inter-node information routing ($\mathrm{dMI}{}_{i,j}$,
$\mathrm{dTE}_{i,j}$) can be lifted to expressions on the inter-community
level ($\mathrm{dMI}{}_{X,Y}$, $\mathrm{dTE_{X,Y}})$ by replacing
node- with community-related quantities (i.e. $\omega_{i}$ with $\Omega_{X}$
or $\gamma_{ik}$ with $\Gamma_{XK}$, etc., Supplementary Corollary~3
and 4). Importantly, this process can be further iterated to networks
of networks, etc. Fig.~\ref{fig: Control} shows examples of information
flow patterns resolved at two scales. The information routing direction
on the larger scale reflects the majority and relative strengths of
IRPs on the finer scale.

\section{Non-Local Information Rerouting via Local Interventions}

\begin{figure*}[!t]
\begin{raggedright}
\centerfloat \includegraphics[scale=0.92]{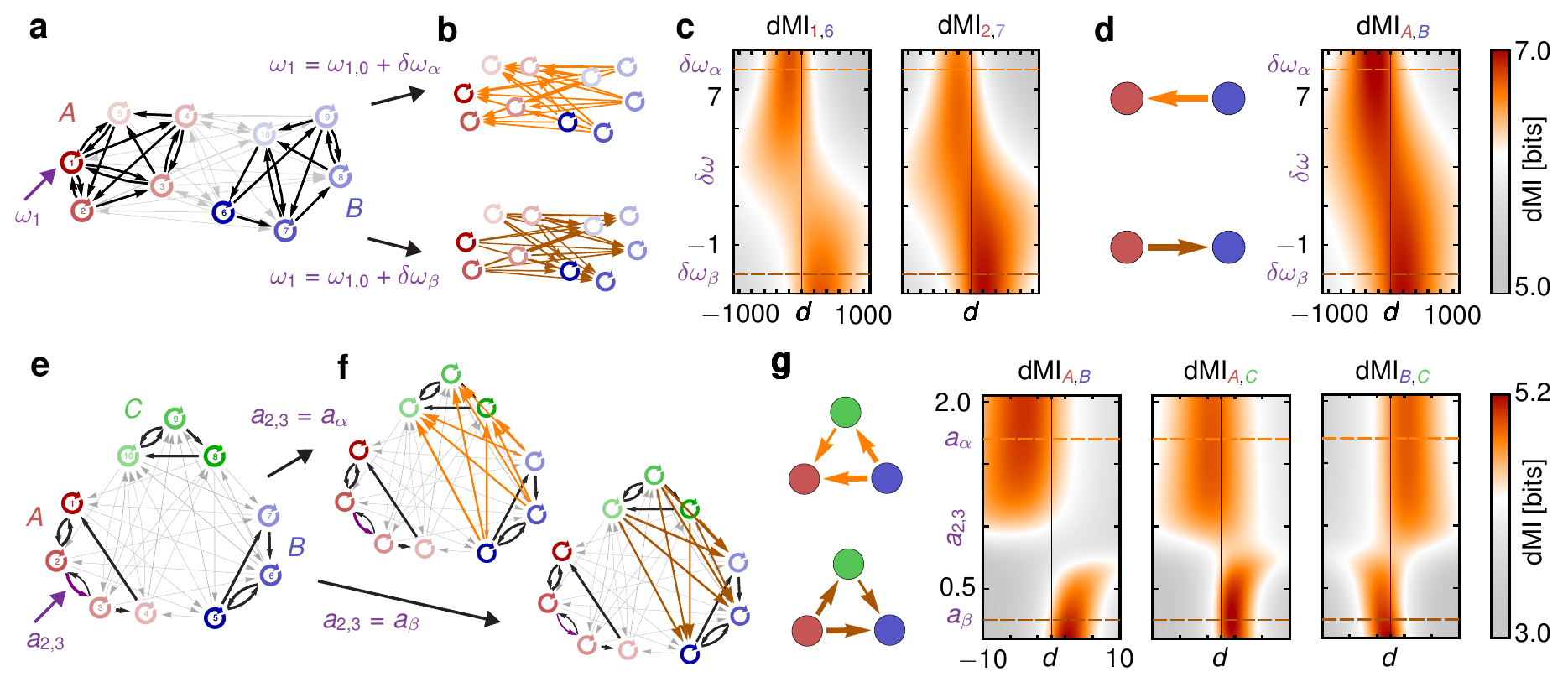}
\par\end{raggedright}

\protect\caption{\label{fig: Control}\textbf{Remote rerouting of information in modular
networks: Local changes trigger global information rerouting. a,}
Network with two coupled communities $A$ and $B$ (red and blue)
of oscillators close to a Hopf bifurcation. Changing the intrinsic
frequency of a single node $i=1$ from $\omega_{1}+\delta\omega_{\alpha}$
to $\omega_{1}+\delta\omega_{\beta}$ induces a collective reorganization
of equilibrium phase differences, that result in \textbf{b, }oppositely
directed information sharing patterns (top vs bottom). \textbf{c,}
dMI between two pairs of nodes from the two different clusters as
a function of the time delay $d$ and frequency change $\delta\omega_{1}$
of oscillator $1$. \textbf{d,} Information flow patterns calculated
from the hierarchically reduced system for the two configurations
in b (left) and as a function of $\delta\omega_{1}$ (right) reflect
the inversion of the IRPs on the finer scale (b,c). \textbf{e,} Network
of three coupled modules of phase oscillators. \textbf{f, }A change
in the connection strength $a_{2,3}$ from $a_{\alpha}$ to $a_{\beta}$
between two nodes ($3_{A}\rightarrow2_{A}$) in sub-network $A$ induces
an inversion of information routing direction between the remote sub-networks
$B$ and $C$. \textbf{g,} Full information routing patterns calculated
form the hierarchical reduced system for $a_{2,3}=a_{\alpha}$ and
$a_{2,3}=a_{\alpha}$ (left) and as a function of $a_{2,3}$ for all
pairs of modules (density plots, right). The transition is not continuous
but rather switch like.}
\end{figure*}

The collective quantities in the system \eqref{eq: reduced dynamical system}
are intricate functions of the network properties at the lower scales.
Intriguingly, the coupling functions $\Gamma_{X,Y}$ not only depend
on the non-local interactions $\gamma_{i_{X}j_{Y}}$ between units
$i_{X}$ of module $X$ and $j_{Y}$ of cluster $Y$ but also on purely
local properties of the individual clusters. In particular, the form
of $\Gamma_{X,Y}$ is a function of the intrinsic local dynamical
states $\mathcal{D}_{X}$ and $\mathcal{D}_{Y}$ of both clusters
as well as the phase response $Z_{X}$ of sub-network $X$ (see Methods
and Supplementary Section 4). Thus IRPs on the entire network level
depend on local community properties. This establishes several generic
mechanisms to globally change information routing in networks via
local changes in modular properties, local connectivity, or via switching
of local dynamical states.

In a network consisting of two sub-networks (Fig.~\ref{fig: Control}a)
the local change of the frequency of a single Hopf-oscillator in sub-network
$A$ induces a non-local inversion of the information routing between
cluster $A$ and $B$ (Fig.~\ref{fig: Control}b-d). In Fig.~\ref{fig: Control}e-f
the direction in which information is routed between two sub-networks
\textbf{$B$} and $C$ of coupled phase oscillators is remotely changed
by increasing the strength of a local link in module $A$. The origin
in both examples is a non-trivial combination of several factors:
The (small) manipulations alter the collective cluster frequency $\Omega_{A}$,
the local dynamical state $\mathcal{D}_{A}$ which in turn change
the collective phase response $Z_{A}$ and the effective noise strength
$\Xi_{A}$ of cluster $A$ (Supplementary Fig.~3). These changes
all contribute to changes in the effective couplings $\Gamma_{X,Y}$
as well as in the inter-cluster phase-locking values $\Delta\Phi_{X,Y}=\Phi_{X}-\Phi_{Y}$.
Taken together this causes the observed inversions in information
routing direction. Interestingly, the transition in information routing
has a switch like dependency on the changed parameter (Fig.~\ref{fig: Control}c,d,g)
promoting digital-like changes of communication modes.

\section{Combinatorial Information Routing Patterns}

\begin{figure}[!t]
\centering{}\textbf{\includegraphics[scale=0.92]{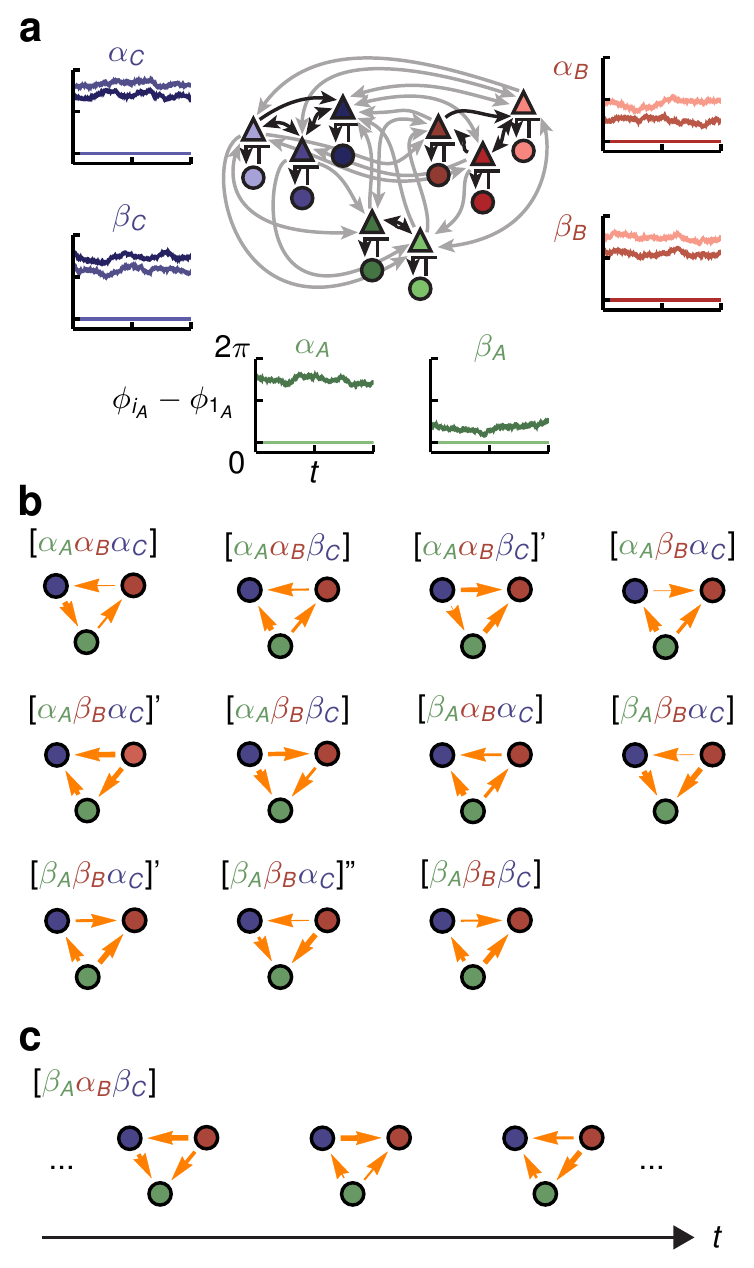}
}\protect\caption{\label{fig:Combinatorics}\textbf{Switching between combinatorially
many information routing patterns.}\emph{ }\textbf{a,} Modular circuit
as in Fig.~\ref{fig:Example_Networks}e. Without inter-module coupling,
each of the $M=3$ communities $X\in\{A,B,C\}$ exhibits multi-stability
between two phase-locked configurations, denoted as states $\alpha_{X}$
and $\beta_{X}$ (insets). \textbf{b,} Information routing patterns
between the hierarchically reduced sub-networks for different combinations
of the local dynamical states $\left[\mathcal{D}_{A}\mathcal{D}_{B}\mathcal{D}_{C}\right]$,
$\mathcal{D}_{X}\in\{\alpha,\beta\}$ that give rise to globally phase-locked
dynamics. Arrows between nodes $X$ and $Y$ indicate strength (line
width) and sign (arrow direction) of the difference in integrated
dTE curves between the nodes. The same local dynamical configuration
($\left[\mathcal{D}_{A}\mathcal{D}_{B}\mathcal{D}_{C}\right]$) can
give rise to more than one globally locked collective state marked
with dashes, i.e. $\left[\mathcal{D}_{A}\mathcal{D}_{B}\mathcal{D}_{C}\right]$,
$\left[\mathcal{D}_{A}\mathcal{D}_{B}\mathcal{D}_{C}\right]'$, $\dots$.
\textbf{c, }The local dynamical state configuration $\left[\beta_{A}\alpha_{B}\beta_{C}\right]$
generates a periodic global dynamical state (cf. Supplementary Figure
8) in which the hierarchically reduced information routing pattern
(graphs) becomes time-dependent (cf.~also Supplementary Section 6).}
\end{figure}

As an alternative to interventions on local properties, also switching
between multi-stable local dynamical states $\mathcal{D}_{X}$ can
induce global information rerouting. In the example in Fig.~\ref{fig:Combinatorics},
each of the $M=3$ modules $X\in\{A,B,C\}$ exhibits $\mathcal{N}_{X}=2$
alternative phase-locked states (labeled $\alpha_{X}$ and $\beta_{X}$,
Supplementary Section 5.1). For sufficiently weak coupling, this local
multi-stability is preserved in the dynamics of the entire modular
network. Consequently each choice of the $\mathcal{N}_{A}\times\mathcal{N}_{B}\times\mathcal{N}_{C}$
possible combinations of ``local'' states gives rise to at least
one network-wide collective state. Certain combinations of local states
can give rise to one or even multiple globally phase-locked states
(e.g. $\left[\alpha_{A}\beta_{B}\alpha_{C}\right]$ in Fig.~\ref{fig:Combinatorics}).
Others support non-phase locked dynamics that gives rise to time-dependent
IRPs (cf.~Fig.~\ref{fig:Combinatorics}c and below). Thus, varying
local dynamical states in a hierarchical network flexibly produces
a combinatorial number $\mbox{\ensuremath{\mathcal{N}}}\ge\prod_{X}\mathcal{N}_{X}$
of different IRPs in the same physical network.

\section{Time-Dependent Information Routing}

General reference states, including periodic or transient dynamics,
are not stationary and hence the expressions for the dMI and dTE become
dependent on time $t$. For example, Fig.~\ref{fig:Combinatorics}c
shows IRPs that undergo cyclic changes due to an underlying periodic
reference state (cf.\ also Supplementary Figure 8a-c). In systems
with a global fixed point systematic displacements to different starting
positions in state space give rise to different stochastic transients
with different and time-dependent IRPs (Supplementary Figure~8d).
Similarly, switching dynamics along heteroclinic orbits constitute
another way of generating specific progressions of reference dynamics.
Thus information 'surfing' on top of non-stationary reference dynamical
configurations naturally yield temporally structured sequences of
IRPs, resolvable also by other measures of instantaneous information
flow, e.g.~\cite{Liang2005,Majda2007,Lizier2008}.

\section{Discussion}

The above results establish a theoretical basis for the emergence
of information routing capabilities in complex networks when signals
are communicated on top of collective reference states. We show how
information sharing (dMI) and transfer (dTE) emerge through the joint
action of local unit features, global interaction topology and choice
of the collective dynamical state. We find that information routing
patterns self-organize according to general principles (cf.~Figs.~\ref{fig: Mechanisms},
\ref{fig: Control}, \ref{fig:Combinatorics}) and can thus be systematically
manipulated. Employing formal identity of our approach at every scale
in oscillatory modular networks (Eq.~\eqref{eq: dynamical system}
vs.~\eqref{eq: reduced dynamical system}) we identify local paradigms
that are capable of regulating information routing at the non-local
level across the whole network (Figs.~\ref{fig: Control}, \ref{fig:Combinatorics}).

\textcolor{black}{In contrast to self-organized technological routing
protocols where local nodes use local routing information to locally
propagate signals, such as in peer-to-peer networks \cite{LuaLim2005},
in the mechanism studied here the information routing modality is
set by the entire network's collective dynamics. This collective reference
state typically evolves on a slower time scale than the information
carrying fluctuations that surf on top of it and is thus different
from signal propagation in cascades \cite{Watts} or avalanches \cite{Beggs}
that dominate on shorter time scales.}

We derived theoretical results based on information sharing and transfer
obtained via delayed mutual information and transfer entropy curves.
Using these abstract measures our results are independent of any particular
implementation of a communication protocol and thus generically demonstrate
how collective dynamics can have a functional role in information
routing. For example, in the network in Fig.~\ref{fig: Mechanisms}
externally injected streams of information are automatically encoded
in fluctuations of the rotation frequency of the individual oscillators.
The injected signals are then transmitted through the network and
decodable from the fluctuating phase velocity of a target unit precisely
along those pathways predicted by the current state-dependent IRP
(Supplementary Section 7).

Our theory is based on a small noise approximation that conditions
the analysis onto a specific underlying dynamical state. In this way
we extracted the precise role of such a reference state for the network's
information routing abilities. For larger signal amplitudes or in
highly recurrent networks in which higher-order interactions can play
an important role the expansion can be carried out systematically
to higher orders using diagrammatic approaches \cite{Kurutcheva}
or numerically to accounting for better accuracy and non-Gaussian
correlations (cf.\ also Supplementary Section 3.4). 

In systems with multi-stable states two signal types need to be discriminated:
those that encode the information to be routed and those that indicate
a switch in the reference dynamics and consequently the IRPs . If
the second type of stimuli is amplified appropriately a switch between
multi-stable states can be induced that moves the network into the
appropriate IRP state for the signals that follow. For example, in
the network of Fig.~\ref{fig: Mechanisms} a switch from state $\alpha$
to $\beta$ can be induced by a strong positive pulse to oscillator
$2$ (and vice versa). If such pulses are part of the input a switch
to the appropriate IRP state will automatically be triggered and the
network auto-regulates its IRP function. More generally a separate
part of the network that effectively filters out relevant signatures
indicating the need for a different IRP could provide such pulses.
Moreover, using the fact that local interventions are capable to switch
IRPs in the network the outcomes of local computations can be used
to trigger changes in the global information routing and thereby enable
context-dependent processing in a self-organized way. 

When information surfs on top of dynamical reference states the control
of IRPs is shifted towards controlling collective network dynamics
making methods from control theory of dynamical systems available
to the control of information routing. For example, changing the interaction
function in coupled oscillators systems \cite{Kiss2007} or providing
control signals to a subset of nodes \cite{Liu2011,Cornelius} are
capable of manipulating the network dynamics. Moreover, switch like
changes (cf.\ Fig.\ \ref{fig: Control}) can be triggered by crossing
bifurcation points and the control of information routing patterns
then gets linked to bifurcation theory of network dynamical systems.

While the mathematical part of our analysis focused on phase signals,
including additional amplitude degrees of freedom into the theoretical
framework can help to explore neural or cell signaling codes that
simultaneously use activity- and phase-based representations to convey
information \cite{Akam2014}. Moreover, separating IRP generation,
e.g. via phase configurations, from actual information transfer, for
instance in amplitude degrees of freedom, might be useful for the
design of systems with a flexible communication function.

The predicted phenomena, including non-local changes of information
routing by local interventions, could be directly experimentally verified
using methods available to date, such as electrochemical arrays \cite{Kiss2007}
or synthetic gene regulatory networks \cite{Stricker2008} (Supplementary
Section 5.3). In addition our results are applicable to the inverse
problem: Unknown network characteristics may be inferred by fitting
theoretical expected dMI and dTE patterns to experimentally observed
data. For example, inferring state-dependent coupling strengths could
further the analysis of neuronal dynamics during context-dependent
processing \cite{Friston2011,Battaglia2012,Canolty,Cole,Kopell2014,ArgwalSommer}.

Modifying inputs, initial conditions or system-intrinsic properties
may well be viable in many biological and artificial systems whose
function requires particular information routing. For instance, on
long time scales, evolutionary pressure may select a particular information
routing pattern by biasing a particular collective state in gene regulatory
and cell signaling networks \cite{Tkacik2008,PurvisLahav,Feillet2014};
on intermediate time scales, local changes in neuronal responses due
to adaptation or varying synaptic coupling strength during learning
processes\textbf{ }\cite{Salazar2012} can impact information routing
paths in entire neuronal circuits; on fast time scales, defined control
inputs to biological networks or engineered communication systems
that switch the underlying collective state, can dynamically modulate
information routing patterns without any physical change to the network.

\textbf{\small{}Methods:}\emph{\small{} Transfer Entropy. }{\small{}The
delayed transfer entropy (dTE)}\emph{\small{} }{\small{}\cite{Schreiber2000}
from a time-series $x_{i}(t)$ to a time-series $x_{j}(t)$ is defined
as}\emph{\small{} }{\small{}
\begin{equation}
\mathrm{dTE}_{i\mapsto j}(d)=\iiint p_{ij,j^{(d)}}\log\left(\frac{p_{ij,j^{\left(d\right)}}p_{j}}{p_{ij}[_{j,j^{(d)}}}\right)\,\mathrm{d}x_{i}\mathrm{d}x_{j}\mathrm{d}x_{j^{\left(d\right)}}\label{eq:TE_general}
\end{equation}
with joint probability $p_{ij,j^{(d)}}=p\left(x_{j}(t+d),x_{i}(t),x_{j}(t)\right)$.
This expression is not invariant under permutation of $i$ and $j$,
implying the directionality of TE. For a more direct comparison with
dMI in Figure~2, we define $\mathrm{dTE_{ij}}(d)$ by $\mathrm{dTE}_{i\rightarrow j}(d)$
for $d>0$ and by $\mathrm{dTE}{}_{j\to i}(-d)$ for $d<0$. }{\small \par}

\emph{\small{}Dynamic Information routing via dynamical states.}{\small{}
For a dynamical system \eqref{eq: general SDE} the reference deterministic
solution $x^{\left(\mathrm{ref}\right)}\left(t+s\right)$ starting
at $x\left(t\right)$ is given by the deterministic flow $x^{\left(\mathrm{ref}\right)}\left(t+s\right)=\mathcal{F}^{\left(\mathrm{ref}\right)}(x\left(t\right),s)$.
The small noise approximation for white noise $\xi$ then yields 
\begin{equation}
p\left(x\left(t+d\right)|x\left(t\right)\right)=\mathcal{N}_{x^{\left(\mathrm{ref}\right)}\left(t+d\right),Q_{d}\left(x\left(t\right)\right)}\left(x\left(t+s\right)\right)\label{eq: general prob}
\end{equation}
where $\mathcal{N}_{x,\Sigma}$ denotes the normal distribution with
mean $x$ and covariance matrix $\Sigma$, $Q_{d}\left(x\right)=\int_{0}^{d}e^{\int_{s}^{d}G\left(r,x\right)dr}\varsigma\varsigma^{\mathrm{T}}e^{\int_{s}^{d}G^{\mathrm{T}}\left(r,x\right)dr}ds$
and $G\left(s,x\right)=Df\left(\mathcal{F}(x,s)\right)$. From this
and the initial distribution $p\left(x\left(t\right)\right)$ the
delayed mutual information and transfer entropy $\mathrm{dMI}_{i,j}\left(d,t\right)$
and $\mathrm{dTE}_{i\rightarrow j}\left(d,t\right)$ are obtained
via \eqref{eq:MI_general} and \eqref{eq:TE_general}. The result
depends on time $t$, lag$d$ and the reference state $x^{\left(\mathrm{ref}\right)}$.}{\small \par}

\emph{\small{}Oscillator Networks.}{\small{} In Fig.~1a, we consider
a network of two coupled biochemical Goodwin oscillators \cite{Goodwin1966,Goldbeter2002}.
Oscillations in the expression levels of the molecular products arise
due to a nonlinear repressive feedback loop in successive transcription,
translation and catalytic reactions. The oscillators are coupled via
mutual repression of the translation process \cite{Wagner2005}. In
addition, in one oscillator changes in concentration of an external
enzyme regulate the speed of degradation of mRNAs, thus affecting
the translation reaction, and, ultimately, the oscillation frequency.
In Fig\@.~\ref{fig:Example_Networks}e, \ref{fig: Mechanisms},
\ref{fig:Combinatorics} we consider networks of Wilson-Cowan type
neural masses (population signals) \cite{Wilson1972}. Each neural
mass intrinsically oscillates due to antagonistic interactions between
local excitatory and inhibitory populations. Different neural masses
interact, within and between communities, via excitatory synapses.
In the generic networks in Fig.~1i and Fig.~3a each unit is modeled
by the normal form of a Hopf-bifurcation in the oscillatory regime
together with linear coupling. Finally, the modular networks analyzed
in Figures~3a and 3b are directly cast as phase-reduced models with
freely chosen coupling functions. See the Supplementary Information
for additional details, model equations and parameters and phase estimation.
}\emph{\small{}}{\small \par}

\emph{\small{}Analytic derivation of the $\mathrm{dMI}$ and $\mathrm{dTE}$
curves.}{\small{} In the small noise expansion \cite{Gardiner2004},
both $\mathrm{dMI}$ and $\mathrm{dTE}$ curves have an analytic approximation:
For stochastic fluctuations around some phase-locked collective state
with constant reference phase offsets $\Delta\phi_{ij}=\phi_{i}-\phi_{j}$
the phases evolve as $\phi_{i}^{(\mathrm{ref})}(t)=\Omega t+\Delta\phi_{i,1}$
in the deterministic limit, where $\Omega=\omega_{i}+\sum_{k}\gamma_{ik}(\Delta\phi_{ik})$
is the collective network frequency and the $\gamma_{ij}(.)$ are
the coupling functions from Eq.~\eqref{eq: dynamical system}. In
presence of noise, the phase dynamics have stochastic components $\phi_{i}^{(\mathrm{fluct})}(t)=\phi_{i}(t)-\phi_{i}^{(\mathrm{ref})}(t)$.
In first order approximation, independent noise inputs $\varsigma_{ij}=\varsigma_{i}\delta_{ij}$
yield coupled Ornstein-Uhlenbeck processes 
\begin{equation}
\frac{\mathrm{d}}{\mathrm{d}t}\phi_{i}^{\left(\mathrm{fluct}\right)}=\sum_{k}g_{ik}\phi_{k}^{\left(\mathrm{fluct}\right)}+\varsigma_{i}\xi_{i}\label{eq:firstnoisexp}
\end{equation}
with linearized, state-dependent couplings given by the Laplacian
matrix entries $g_{ij}=-\gamma_{ij}'\left(\Delta\phi_{ij}\right)$
and $g_{ii}=\sum_{k}\gamma_{ik}'\left(\Delta\phi_{ik}\right)$. The
analytic solution to the stochastic equations \eqref{eq:firstnoisexp}
provides an estimate of the probability distributions, $P_{i}$, $P_{ij^{(d)}}$
and $P_{ij,j^{\left(d\right)}}$. Via \eqref{eq:MI_general} this
results in a prediction for $\mathrm{dMI}_{ij}(d)$, Eq.~\eqref{eq:MI-1},
as a function of the matrix elements $k_{ij^{(d)}}$ specifying the
inverse variance of a von Mises distribution ansatz for $P_{ij^{(d)}}$.
Similarly via \eqref{eq:TE_general} an expression for $\mathrm{dTE}_{i\rightarrow j}(d)$
is obtained. For the dependency of $k_{ij^{(d)}}$, and $\mathrm{dTE}_{i\rightarrow j}(d)$
on network parameters and further details, see the derivation of the
Theorems 1 and 2 in the Supplementary Information.}{\small \par}

\emph{\small{}Time scale for information sharing.}{\small{} For a
network of two oscillators as in Fig.~\eqref{fig: Mechanisms} with
linearized coupling strengths $g_{1}^{\left[\alpha\right]}$ and $g_{2}^{\left[\alpha\right]}$
and $g_{1}^{\left[\alpha\right]}<g_{2}^{\left[\alpha\right]}$, maximizing
$\mathrm{dMI}_{12}\left(d\right)$ (see Supplementary Information
for full analytic expressions of $\mathrm{dMI}$ and $\mathrm{dTE}$
in two oscillator networks) yields}{\small \par}

{\small{}
\begin{equation}
d^{*}=\left(g_{1}^{\left[\alpha\right]}+g_{2}^{\left[\alpha\right]}\right)^{-1}\log\left(\frac{1}{2}\left(1+\left(\frac{g_{2}^{\left[\alpha\right]}}{g_{1}^{\left[\alpha\right]}}\right)^{2}\right)\right)\label{eq: time scale}
\end{equation}
}{\small \par}

\emph{\small{}Collective phase reduction. }{\small{}Suppose that each
node $i=i_{X}$ belongs to a specific network module $X$ out of $M\leq N$
non-overlapping modules of a network. Then equation \eqref{eq: dynamical system}
can be simplified to \eqref{eq: reduced dynamical system} under the
assumption that in the absence of noise every community $X$ has a
stable internally phase-locked state $\phi_{i_{X}}^{(\mathrm{ref})}(t)=\Phi_{X}(t)+\Delta\phi_{i_{X}}$,
where $\Delta\phi_{i_{X}}$ are constant phase offsets of individual
nodes $i_{X}$. Every community can then be regarded as a single meta-oscillator
with a collective phase $\Phi_{X}(t)$ and a collective frequency
$\Omega_{X}=\omega_{i_{X}}+\sum_{j_{X}}\gamma_{i_{X}j_{X}}(\Delta\phi_{i_{X}}-\Delta\phi_{j_{X}})$.
The vector components of the collective phase response $Z_{X}$, the
effective couplings $\Gamma_{XY}$ and the noise parameters $\Sigma_{XY}$
and $\Xi_{X}$ are obtained through collective phase reduction and
depend on the respective quantities ($\omega_{i_{X}},\gamma_{i_{X}j_{X}},\ldots$)
on the single-unit scale (see Supplementary Section 4 for a full derivation).}\\
{\small \par}

\textbf{Acknowledgements:} We thank T. Geisel for valuable discussions.
Partially supported by the Federal Ministry for Education and Research
(BMBF) under grants no. 01GQ1005B {[}CK, DB, MT{]} and 03SF0472E {[}MT{]},
by the NVIDIA Corp., Santa Clara, USA {[}MT{]}, a grant by the Max
Planck Society {[}MT{]}, by the FP7 Marie Curie career development
fellowship IEF 330792 (DynViB) {[}DB{]} and an independent postdoctoral
fellowship by the Rockefeller University, New York, USA {[}CK{]}.\\

\textbf{Author contributions:} All authors designed research. C.K.
derived the theoretical results, developed analysis tools and carried
out the numerical experiments. All authors analyzed and interpreted
the results and wrote the manuscript.\\

\textbf{Additional information:} The authors declare no competing
financial interests. Supplementary information accompanies this paper.
Correspondence and requests should be addressed to CK (ckirst@rockefeller.edu).\\

\end{document}